\documentclass[prl,twocolumn,showpacs,showkeys,%
  preprintnumbers,amsmath,amssymb]{revtex4}

\usepackage{graphicx,color}
\usepackage{dcolumn}
\usepackage{bm}

\def\beq{\begin{eqnarray}}
\def\eeq{\end{eqnarray}}

\begin{document}

\preprint{SI-HEP-2008-01, TUM-HEP-680/08}

\title{Large Top Mass and
Non-Linear Representation of Flavour Symmetry}

\author{Thorsten Feldmann}%
\altaffiliation[Also at: ]{Technische Universit\"at M\"unchen, Physik Department, 85747 Garching, Germany}%
\email{feldmann@hep.physik.uni-siegen.de}
\author{Thomas Mannel}%
 \email{mannel@hep.physik.uni-siegen.de}

\affiliation{%
Theoretische Physik I, Fachbereich Physik, Universit\"at Siegen, 
  57068 Siegen, Germany
}%

\date{January 11, 2008}

\begin{abstract}

 We consider an effective theory (ET) approach to flavour-violating 
 processes beyond the Standard Model (SM), where the breaking of 
 flavour symmetry is described by spurion fields whose low-energy
 vacuum expectation values are identified with the SM Yukawa couplings. 
 Insisting on canonical mass dimensions for
 the spurion fields, the large top-quark
 Yukawa coupling also implies a large expectation value
 for the associated spurion, which breaks part of the
 flavour symmetry already at the UV scale $\Lambda$ of the ET.
 Below that scale, flavour symmetry in the ET is represented in a
 non-linear way by introducing Goldstone modes
 for the partly broken flavour symmetry and spurion fields 
 transforming under the residual symmetry. As a result, the
 dominance of certain flavour structures in rare quark decays 
 can be understood in terms of
 the $1/\Lambda$ expansion in the ET.
 We also discuss the generalization
 to 2-Higgs--doublet models with large $\tan\beta$.
\end{abstract}

\maketitle




Understanding the origin behind the existence of three quark and 
lepton families is one of the biggest challenges in contemporary particle 
physics. While many arguments have been given that the explanation
of the electroweak symmetry breaking in the SM requires new physics 
at the TeV scale, the recent experiments at $B$-meson factories seem
to indicate that quark flavour violation beyond the known sources
in the Standard Model (SM) happens at much higher scales.
For this reason it has been proposed to consider the SM
and its potential extensions as an effective theory (ET), where the 
concept of quark (and also lepton) flavour is introduced by a 
symmetry principle that guarantees minimal flavour violation (MFV), 
i.e.\ any new source for quark flavour transitions should be induced by the 
Yukawa coupling matrices for up- and down-type quarks of the SM
\cite{D'Ambrosio:2002ex}. 
As we will explain below, in the original formulation of MFV 
the large top-quark Yukawa coupling distorts the conventional
power-counting in the ET. In this letter, we show how the special
role of the top quark can be implemented into the ET approach
in a natural way, which also leads to some new insights
into the structure of MFV.

As is well known \cite{Chivukula:1987py}, 
the maximal quark flavour symmetry group
commuting with the SM gauge symmetry is
\begin{equation} \label{FlavourGroup}
G_F = SU(3)_{Q_L} \times SU(3)_{U_R} \times SU(3)_{D_R} \,.
\end{equation}
Additional $U(1)$ factors, which become relevant for the
discussion of 2-Higgs-doublet models \cite{D'Ambrosio:2002ex}, 
can be identified with baryon number, hyper-charge and the 
Peccei-Quinn symmetry \cite{Peccei:1977ur}.
The quarks are grouped into left-handed doublets and right-handed 
singlets with respect to weak isospin. Under the flavour symmetry
$G_F$, they transform as
\begin{equation}
Q_L  \sim (3,1,1) \,, \quad
U_R  \sim (1,3,1) \,, \quad
D_R  \sim (1,1,3) \,,
\end{equation}
while the SM gauge and Higgs fields are singlets with respect to
$G_F$.

Considering the SM as the leading (dimension-4) part of an effective
theory, the breaking of the flavour group can be achieved by  
introducing spurion fields \cite{D'Ambrosio:2002ex}, transforming 
under $G_F$ as
\begin{equation} 
 Y_U  \sim (3,\bar{3},1) \,, \qquad 
Y_D \sim (3,1,\bar{3}) \,.
\end{equation}
If we assume the spurions to have canonical mass dimension one, 
the Yukawa couplings in the SM, which couple left- and right-handed
quark fields to the Higgs doublet, already stem from 
dimension-five operators,
\begin{equation}
- {\cal L}_{\rm Yuk} = \frac{1}{\Lambda} (\bar{Q}_L \tilde{\phi}) Y_U U_R + 
 \frac{1}{\Lambda}( \bar{Q}_L \phi) Y_D D_R + {\rm h.c.} 
\end{equation}
where $\phi$ is the usual Higgs doublet, 
$\tilde{\phi}$ its charge conjugate, and $\Lambda$ denotes some 
high-energy scale, $\Lambda \gg M_W$.

If the spurion fields $Y_U$ and $Y_D$ acquire vacuum expectation values
at an intermediate scale $\Lambda' \ll \Lambda$, one could easily explain the 
smallness of the $u,c,d,s,b$ quark masses, while the top quark would
not fit into this scheme and requires $\Lambda' \sim \Lambda$. In
other words, the top-quark Yukawa coupling should rather be described
by a dimension-4 operator in the ET below $\Lambda$. 
In the usual (linear) formulation of MFV
\cite{D'Ambrosio:2002ex}, 
the unusually large value of $y_t$ thus distorts the conventional 
power-counting of the spurion analysis.
Therefore, it would be desirable to single out the role of the top
quark more explicitly. To this end,
we are going to construct a non-linear representation of the flavour symmetry
group in which the concept of MFV
can be embedded in an economic way.

%
%
%
Starting from the flavour group $G_F$, which is 
considered a good symmetry at scales above $\Lambda$,
we now assume that at the scale $\Lambda$
the spurion $Y_U$ acquires a vacuum expectation value (VEV)
 \footnote{In the following,
hatted quantities denote dimensionless Yukawa matrices, whereas
the corresponding unhatted fields are spurions with canonical
mass dimension.},
\begin{equation}
\langle \hat Y_U \rangle \equiv 
\langle \frac{Y_U}{\Lambda} \rangle = \left( \begin{array}{ccc} 
0 & 0 & 0 \\
0 & 0 & 0 \\
0 & 0 & y_t \end{array} \right) \,,
\label{yt}
\end{equation}  
which breaks the original flavour group $G_F$ down to 
\begin{equation}
G_F^\prime = SU(2)_{Q_L} \times SU(2)_{U_R} \times SU(3)_{D_R} \times U(1)_T
\,.
\end{equation}
The reduced flavour group $G_F^\prime$ corresponds to the limit where
all Yukawa couplings except for $y_t$ and all off-diagonal entries
in the CKM matrix are neglected.
The 7 unbroken generators of the subgroup
$SU(2)_{Q_L} \times SU(2)_{U_R} \times U(1)_T$ of $G_F^\prime$
are identified as $T_{Q_L,U_R}^{a=1,2,3}$,
and a charge operator,
\begin{equation}
 Q_T = B - \frac{2}{\sqrt 3} \, T^8_{Q_L=U_R} \,,
\label{QT}
\end{equation}
which acts on the 3$^{\rm rd}$-family component of $Q_L$ and $U_R$,
only (here $B$ denotes baryon number).
The ``Goldstone'' modes, corresponding to the 9 broken generators of $G_F/G_F'$,
are written in the standard parameterization~\cite{Coleman:1969sm}, 
\begin{equation}
{\cal U}(\Pi_X) = \exp \left(\frac{i}{\Lambda} \, \sum_{a=4}^8 \,
 T_X^a \, \Pi_X^a \right) \,, 
\label{UX}
\end{equation}
with $X=Q_L,U_R$ and the constraint $\Pi_{Q_L}^8 = - \Pi_{U_R}^8$ (the case $\Pi_{Q_L}^8=\Pi_{U_R}^8$ is
included in (\ref{QT})). 

The remaining degrees of freedom in the Yukawa matrix
$\hat Y_U$ can be parameterized as 
\begin{equation}
 \hat Y_U = {\cal U}(\Pi_{Q_L}) \,
\left( \begin{array}{cc} 
Y_U^{(2)}/\Lambda & \begin{array}{c} 0 \\ 0 \end{array} \\
\begin{array}{cc} 0 & 0 \end{array} & y_t
\end{array}
\right)\, {\cal U}^\dagger(\Pi_{Q_R}) \,.
\label{YU2}
\end{equation}
where the residual spurion $Y_U^{(2)}$ now has
canonical mass dimension one, and transforms
as $(2,2,1)_0$ under $G_F'$.
It is supposed to
develop its VEV at a lower scale, $\Lambda'\ll \Lambda$,
which breaks $SU(2)_{Q_L} \times SU(2)_{Q_R}$.
The 8 real parameters of
the complex $2\times 2$ matrix $Y_U^{(2)}$, the top-Yukawa coupling $y_t$,
and the 9 Goldstone degrees of freedom add up to 18 real parameters 
describing the complex $3\times 3$ matrix $Y_U$. 

The matrices ${\cal U}(\Pi_{Q_L})$ and ${\cal U}(\Pi_{U_R})$
can be further reduced, according to
\begin{align}
  ({\cal U}(\Pi_{X})_{ij} =& (\Xi_{X})_i \, \delta_{j3} +
   \sum_{k=1,2} \, ({\cal U}_{X}^{(2)})_{ik} \, \delta_{kj} \,,
\label{reduce}
\end{align}
with the individual components transforming as
3-vectors and $3\times 2$ matrices, respectively,
\begin{align}
 \Xi_X(\Pi_X)  \to {} & \Xi_X(\Pi_X') = V_X \, \Xi_X(\Pi_X) \,,
\nonumber \\[0.2em]
 {\cal U}_{X}^{(2)}(\Pi_X)  \to {} &  {\cal U}_{X}^{(2)}(\Pi_X') 
= V_{X} \, {\cal U}_X^{(2)}(\Pi_X) \, V_{X}^{(2)\dagger} \,,
\end{align}
with $V_X \in SU(3)_X$ and $V_X^{(2)} \in SU(2)_X$.
Notice that in general, 
$
 \Pi_{X}'
$
is a complicated non-linear function of $\Pi_{X}$.
However, on the subgroup $G_F'$ of $G_F$, the $\Pi_{X}$ transform
linearly, according to
\begin{equation}
(\Pi_X')^a = {\rm tr} \left[ T_{X}^a \, V_{X}^{(2)} \, \Pi_{X} \, V_{X}^{(2)\dagger} \right] \,.
\end{equation}
The unitarity of ${\cal U}$ implies the following relations
\begin{equation}
 \sum_{j=1,2} ({\cal U}_X^{(2)})_{ij} \, ({\cal U}_X^{(2) \dagger})_{jk} 
    + (\Xi_X)_i \, (\Xi_X^\dagger)_k = \delta_{ik} \,,
\label{unit1}
\end{equation}
and
\begin{align}
&
 {\cal U}_X^{(2)\dagger} \, {\cal U}_X^{(2)} = {\boldsymbol{1} } \,, \ \quad
\Xi_X^\dagger \, \Xi_X = 1 \,,
\nonumber
 \\[0.2em]
&
 {\cal U}_X^{(2)\dagger} \, \Xi_X =
        \Xi_X^\dagger \, {\cal U}_X^{(2)} = 0 \,.
\label{unit2}
\end{align}

The SM Yukawa term for up-type quarks in the non-linear
representation can then be written in manifestly invariant form as
\begin{align}
- {\cal L}_{\rm yuk}^{(u)} = {} &
 y_t \,   
(\bar Q_L \, \Xi_{Q_L}) \, \tilde \phi \, (\Xi_{U_R}^\dagger  U_R)
\nonumber \\[0.2em]
& {} + \frac{1}{\Lambda} \,
(\bar Q_L \, {\cal U}_{Q_L}^{(2)} \ \tilde\phi \ Y_U^{(2)} \, 
                  {\cal U}_{U_R}^{(2)\dagger} \, U_R) +\mbox{h.c.}
\label{yuk_u}
\end{align}
Similarly, the Yukawa terms for down-type quarks
are obtained from
\begin{align}
- {\cal L}_{\rm yuk}^{(d)} = {} &
\frac{1}{\Lambda} \, (\bar Q_L \, \Xi_{Q_L}) \, \phi \,
  (\xi_{D_R}^\dagger  D_R)
\nonumber \\[0.2em]
& {} + \frac{1}{\Lambda} \,
(\bar Q_L \, {\cal U}_{Q_L}^{(2)} \ \phi \ Y_D^{(2)} \,
 D_R)
+\mbox{h.c.}
\label{yuk_d}
\end{align}
where we have identified two
irreducible spurions of $G_F'$ in addition to $Y_U^{(2)}$ in (\ref{YU2}),
\begin{align}
 \xi_{D_R}^\dagger  = {} 
  & \Xi_{Q_L}^\dagger \, \hat Y_D \, \Lambda \  \sim \ (1,1,\bar 3)_{+1} \,,
\nonumber \\[0.2em]
 Y_D^{(2)}  = {} &
 {\cal U}_{Q_L}^{(2)\dagger} \, \hat Y_D \, \Lambda \ \sim \ (2,1,\bar 3)_0 \,.
\end{align}
Again, the new spurion fields have canonical mass dimension,
and assume VEVs of the order $\Lambda'\ll \Lambda$.
Therefore, by construction, in the limit $\Lambda \to \infty$,
only the dimension-4 term for the top Yukawa coupling 
(first line in (\ref{yuk_u})) survives,
while the dimension-5 operators in (\ref{yuk_u},\ref{yuk_d}),
related to the breaking of $G_F'$, vanish.

Noticing that the spurion matrices $Y_U^{(2)}$ and $Y_D$ can
-- as usual -- be diagonalized by bi-unitary transformations,
\begin{align}
 & Y_U^{(2)} = V_{u_L}^{(2)} \, Y_U^{\rm (2) diag} \, V_{u_R}^{(2)} \,,
\cr
 & Y_D = V_{d_L} \, Y_D^{\rm diag} \, V_{d_R} \,,
\label{diag1}
\end{align}
the CKM matrix in the SM is identified as
\begin{equation}
  V_{\rm CKM} = 
 \left( \begin{array}{c}
    V_{u_L}^{(2)\dagger} \, {\cal U}_{Q_L}^{(2)\dagger} \\ 
  \Xi_{Q_L}^\dagger \end{array}\right) \, V_{d_L} \,.
\label{VCKM}
\end{equation}


The construction of MFV operators
in the ET is now straightforward.
Let us consider the decay $b \to s\gamma$
as an example. 
In the SM, it is generated by 
flavour-changing neutral currents, arising from
loop diagrams with charged gauge bosons and
up-type quarks.
In the ET, new physics can contribute via higher-dimensional operators
above the electro-weak scale. An example is given by 
\begin{equation}
 {\cal O}_{\rm eff} = \frac{1}{\Lambda^3} \,
 (\bar Q_L  \Xi_{Q_L} \, \phi \, \sigma_{\mu\nu} \
  \xi_{D_R}^\dagger  D_R) \, F^{\mu\nu} + {\rm h.c.}
\label{Oeff}
\end{equation}
which is manifestly invariant under gauge and flavour transformations
(a second flavour structure involving $Y_D^{(2)}$ is related to (\ref{Oeff})
and a flavour-diagonal operator by the unitarity relation (\ref{unit1})).
After changing to the mass basis, using
(\ref{diag1},\ref{VCKM}),
we obtain
\begin{equation}
\left( V_{d_L}^\dagger \, \Xi_{Q_L} \, \xi_{D_R}^\dagger  V_{d_R} \right)_{ij}
=  V_{ti}^* V_{tj} \, (y_d)_j \, \Lambda \,.
\label{s1}
\end{equation}

This may be compared with the standard (linear) formulation of MFV 
\cite{D'Ambrosio:2002ex}, where the effective operator would be written
as
\begin{equation}
 {\cal O}_{\rm eff} = \frac{1}{\Lambda^2} \,
 (\bar Q_L \, \hat Y_U \hat Y_U^\dagger  \, \phi \, \sigma_{\mu\nu} \
  \hat Y_D \,  D_R) \, F^{\mu\nu} + {\rm h.c.}
\label{Oeff2}
\end{equation}
with the resulting flavour
coefficients proportional to
\begin{equation}
  \sum_{k=u,c,t} V_{ki}^* \, (y_k)^2  \, V_{kj} \, (y_d)_j  \,,
\label{s2}
\end{equation}
which coincides with (\ref{s1}) up to terms of order $y_c^2/y_t^2$.
Inserting the Higgs VEV into (\ref{Oeff}) and projecting on
$b \to s$ transitions, we obtain contributions to the low-energy
weak effective Hamiltonian \cite{Buchalla:1995vs} with
flavour coefficients 
\begin{align}
 \frac{m_W^2}{\Lambda^2} \, m_b \, V_{ts}^* V_{tb} & \qquad 
  \mbox{(for $b_R \to s_L$ transitions)}
\nonumber \\[0.2em]
 \frac{m_W^2}{\Lambda^2} \, m_s \, V_{tb}^* V_{ts} & \qquad 
  \mbox{(for $b_L \to s_R$ transitions)} 
\end{align}
Our example exhibits the usual advantage of the non-linear
representation for the power-counting in ETs 
with spontaneously broken global symmetries: 
The dominance of the top-quark contribution in (\ref{s2})
is a manifest consequence of the $1/\Lambda$ expansion. 
Also the chiral suppression factors $m_b,m_s$
can be traced back to the additional power of $1/\Lambda$
in (\ref{Oeff}) compared to (\ref{Oeff2}), i.e.\ in
the non-linear representation NP contributions to $b\to s\gamma$
are related to dimension-7 operators in the ET
above the electro-weak scale.


Quark bi-linears with other chirality structures can be made invariant
under the flavour group in a similar fashion as in the above example.
The possible structures fall into three classes, which can be constructed
in terms of fundamental building blocks which transform as singlets
under $SU(2)_{U_R} \times SU(3)_{D_R}$. The first class contains
all possible combinations of the form
\begin{align}
 \left( \begin{array}{l}
         \bar Q_L \, \Xi_{Q_L} \\
         \bar U_R \, \Xi_{U_R} \\
         \bar D_R \, \xi_{D_R}
        \end{array} \right) 
\otimes 
 \left( \begin{array}{r}
         \Xi_{Q_L}^\dagger Q_L  \\
         \Xi_{U_R}^\dagger U_R  \\
         \xi_{D_R}^\dagger D_R 
        \end{array} \right) \,.
\end{align}
The second class is constructed from
\begin{align}
 \left( \begin{array}{l}
         \bar Q_L \, {\cal U}_{Q_L}^{(2)} \\
         \bar U_R \, {\cal U}_{U_R}^{(2)} \, Y_U^{(2)^\dagger} \\
         \bar D_R \, Y_D^{(2)\dagger}
        \end{array} \right) 
\otimes 
{\cal P} 
\otimes
 \left( \begin{array}{r}
         {\cal U}_{Q_L}^{(2)\dagger} Q_L  \\
         Y_U^{(2)} \, {\cal U}_{U_R}^{(2)\dagger} U_R  \\
         Y_D^{(2)} \, D_R 
        \end{array} \right) \,,
\end{align}
where ${\cal P}$ is an arbitrary polynomial
of $Y_U^{(2)} \, Y_U^{(2)\dagger}$ and $Y_D^{(2)} \, Y_D^{(2)\dagger}$,
which transforms as $(1,1,1)_0+(3,1,1)_0$.
Finally, the third class involves
\begin{align}
 \left( \begin{array}{l}
         \bar Q_L \, {\cal U}_{Q_L}^{(2)} \\
         \bar U_R \, {\cal U}_{U_R}^{(2)} \, Y_U^{(2)^\dagger} \\
         \bar D_R \, Y_D^{(2)\dagger}
        \end{array} \right) 
\otimes 
{\cal P} 
\cdot Y_D^{(2)} \xi_{D_R} 
\otimes
\left( \begin{array}{r}
         \Xi_{Q_L}^\dagger Q_L  \\
         \Xi_{U_R}^\dagger U_R  \\
         \xi_{D_R}^\dagger D_R 
        \end{array} \right) +\mbox{h.c.}
\end{align}

%
%
%

Our discussion can be generalized to models with an extended Higgs
sector.
Of particular interest is the
2-Higgs--doublet models (2HDM) of type~II \cite{Gunion:1984yn}, 
where the small value of the bottom quark mass 
can be related to a large ratio of
Higgs VEVs $\tan\beta =v_u/v_d$, such that
$$
  y_b = m_b/v_d \sim y_t = m_t/v_u \sim 1\,.
$$
The non-linear representation of the flavour symmetry for
this case can be achieved by considering a VEV
for the spurion field $\xi_{D_R}^\dagger$,
\begin{equation}
 \langle \xi_{D_R}^\dagger \rangle = (0,0,\tilde y_b)\,\Lambda \,,
\end{equation}
in addition to (\ref{yt}).
It breaks a sub-group of $G_F'$,
\begin{equation}
 SU(3)_{D_R} \times U(1)_T \to SU(2)_{D_R} \times U(1)_{\rm III} \,,
\end{equation}
such that the unbroken flavour group is now given by
\begin{equation}
 G_F''= SU(2)_{Q_L} \times SU(2)_{U_R} \times SU(2)_{D_R} \times U(1)_{\rm III} \,.
\end{equation}
(Alternatively, we could have assumed a large VEV for the spurion field $Y_D^{(2)}$.
However, in this case one would have off-diagonal CKM elements of order 1 in
contrast to observation.)
The 10 unbroken generators of $G_F''$
are $T_{Q_L,U_R,D_R}^{a=1,2,3}$ and a charge
operator for the 3$^{\rm rd}$ generation,
\begin{equation}
 Q_{\rm III} = B - \frac{2}{\sqrt 3} \, T^8_{Q_L=U_R=D_R} \,.
\end{equation}
The additional 5 Goldstone bosons associated to the breaking of $G_F' \to G_F''$
now appear in the non-linear representation of the spurion field 
\begin{equation}
 \xi_{D_R}^\dagger = \langle \xi_{D_R}^\dagger \rangle \, {\cal U}^\dagger(\Pi_{D_R}) 
= \tilde y_b \, {\Xi}_{D_R}^\dagger \, \Lambda \,,
\end{equation}
where ${\cal U}(\Pi_{D_R})$ and $\Xi_{D_R}$ are defined analogously to (\ref{UX},\ref{reduce}).
The reduction of the remaining spurion field $Y_D^{(2)}$ in terms of
irreducible representations of $G_F''$ is given by
\begin{align}
 \tilde Y_D^{(2)} \equiv {} &   Y_D^{(2)} \, {\cal U}^{(2)}_{D_R} \sim (2,1,2)_0 \,, 
\nonumber \\[0.2em]
 \chi_{L} \equiv {} &   Y_D^{(2)} \, \Xi_{D_R} \sim (2,1,1)_{-1} \,.
\end{align}
In this way, the down-type Yukawa matrix is now parameterized as
\begin{equation}
  \hat Y_{D} = 
{\cal U}(\Pi_{Q_L}) \,
\left( \begin{array}{cc} 
\tilde Y_{D}^{(2)}/\Lambda & 
\chi_L /\Lambda
\\
\begin{array}{cc} 0 & 0 \end{array} 
& \tilde y_{b}
\end{array}
\right)\, {\cal U}^\dagger(\Pi_{D_R}) \,,
\label{Y2HDM}
\end{equation}
which is to be diagonalized as
\begin{equation}
\hat Y_D =
V_{d_L} \left( \begin{array}{ccc} y_d & 0 & 0 \\ 0 & y_s & 0 \\ 0 &0 & y_b
               \end{array} \right) V_{d_R}^\dagger \,,
\label{diagneu}
\end{equation}
with two small eigenvalues $y_{d,s} = {\cal O}(\Lambda'/\Lambda)$.
In particular, we have the relations
\begin{align}
& \left( {\cal U}^\dagger(\Pi_{Q_L}) \, \hat Y_D \, V_{D_R} \right)_{3j} 
=  \tilde y_b \left( \Xi_{D_R}^\dagger V_{d_R} \right)_j
\cr
= {} &( \Xi_{Q_L}^\dagger V_{d_L})_{3j} \, (y_d)_j 
= V_{tb} \, y_b \, \delta_{3j} + {\cal O}(\Lambda'/\Lambda) \,,
\label{yb}
\end{align}
and
\begin{align}
 |\tilde y_b|^2 = \sum_{j=1}^3 \, V_{tj} \,  (y_d^2)_j \, V_{tj}^*
 = y_b^2 +  {\cal O}((\Lambda'/\Lambda)^2) \,.
\end{align}

Considering again the example $b \to s\gamma$ in the case of 
large $\tan\beta$, the analogue of (\ref{Oeff}) is now given by
a dimension-6 operator,
\begin{equation}
 {\cal O}_{\rm eff} 
= \frac{1}{\Lambda^2} \,
 (\bar Q_L  \Xi_{Q_L} \, \phi_D \, \sigma_{\mu\nu} \
  \Xi_{D_R}^\dagger D_R) \, F^{\mu\nu} + {\rm h.c.}
\label{Oeffneu}
\end{equation}
After changing to the mass basis, using
(\ref{VCKM},\ref{yb}),
we obtain
\begin{equation}  
\left(V_{d_L}^\dagger  \Xi_{Q_L} \, \Xi_{D_R}^\dagger  V_{d_R} \right)_{ij}
= V_{ti}^* \, V_{tj} \, (y_d)_j/\tilde y_b
\simeq V_{ti}^* \, V_{tb} \, \delta_{3j} \,,
\end{equation}
which projects out the leading chirality structure in (\ref{s1})
for a large Yukawa coupling of the bottom quark~\footnote{Notice that for $b \to s\gamma$ 
the VEV for $\phi_D$ in (\ref{Oeffneu}) provides an additional suppression factor
$v_d/v = \cos\beta$.}.

The set of invariant quark bi-linears
in the 2HDM with large $\tan\beta$
is constructed from building blocks which are singlets under
$SU(2)_{U_R} \times SU(2)_{D_R}$, namely~\footnote{
We do not discuss the issue of broken Peccei-Quinn symmetry~\cite{Peccei:1977ur} here.
A detailed discussion can be found in \cite{D'Ambrosio:2002ex}.}
\begin{align}
 \left( \begin{array}{l}
         \bar Q_L \, \Xi_{Q_L} \\[0.3em]
         \bar U_R \, \Xi_{U_R} \\[0.3em]
         \bar D_R \, \Xi_{D_R}
        \end{array} \right) 
 \otimes
 \left( \begin{array}{r}
         \Xi_{Q_L}^\dagger Q_L  \\[0.2em]
         \Xi_{U_R}^\dagger U_R  \\[0.2em]
         \Xi_{D_R}^\dagger D_R 
        \end{array} \right) \,,
\end{align}
and
\begin{align}
 \left( \begin{array}{l}
         \bar Q_L \, {\cal U}_{Q_L}^{(2)} \\[0.2em]
         \bar U_R \, {\cal U}_{U_R}^{(2)} \\[0.2em]
         \bar D_R \, {\cal U}_{D_R}^{(2)}
        \end{array} \right) 
\otimes 
 {\cal P}'
\otimes
 \left( \begin{array}{r}
         {\cal U}_{Q_L}^{(2)\dagger} Q_L  \\[0.2em]
         {\cal U}_{U_R}^{(2)\dagger} U_R  \\[0.2em]
         {\cal U}_{D_R}^{(2)\dagger} D_R 
        \end{array} \right) \,,
\end{align}
and
\begin{align}
 \left( \begin{array}{l}
         \bar Q_L \, {\cal U}_{Q_L}^{(2)} \\[0.2em]
         \bar U_R \, {\cal U}_{U_R}^{(2)} \\[0.2em]
         \bar D_R \, {\cal U}_{D_R}^{(2)}
        \end{array} \right) 
\otimes 
 {\cal P}' \cdot \chi_L
\otimes
 \left( \begin{array}{r}
         \Xi_{Q_L}^\dagger Q_L  \\[0.3em]
         \Xi_{U_R}^\dagger U_R  \\[0.3em]
         \Xi_{D_R}^\dagger D_R 
        \end{array} \right) \,,
\end{align}
with ${\cal P}'$ being an arbitrary polynomial of
$Y_U^{(2)} \, Y_U^{(2)^\dagger}$,
$\tilde Y_D^{(2)} \, \tilde Y_D^{(2)^\dagger}$,
and $\chi_L \, \chi_L^\dagger$.


In summary, we have constructed an effective theory for
flavour transitions beyond the SM, where the original
(global) flavour symmetry $G_F$ of the SM gauge sector
is considered to be 
spontaneously broken by the large top -- and, in the
case of 2HDM with large $\tan \beta$, also bottom --
Yukawa coupling. The associated Goldstone modes can be used to define
a non-linear representation of flavour symmetry, where
the concept of minimial flavour violation can be embedded 
by introducing spurion fields with canonical mass dimension.
As a consequence, the dominance of certain flavour structures
in rare quark decays -- in particular, the top-quark dominance
in flavour-changing neutral currents with down-type quarks --
is a direct consequence of the $1/\Lambda$ expansion in the
effective theory. The dynamical interpretation of the Goldstone
modes as well as possible extensions beyond MFV (see, for instance, \cite{Feldmann:2006jk})
are left for future studies.

\begin{acknowledgements}
We thank Nima Arkani-Hamed for discussions on
the non-linear realization of flavor symmetries and related
issues independently under exploration by Gilad Perez and 
Tomer Volansky.
This work is supported by the German Ministry of Research
(BMBF, contract No.~05HT6PSA). T.F.\  has been supported
by the Cluster of Excellence ``Origin and Structure of the Universe''.

\end{acknowledgements}

\end{document}